# Closed-loop experiments and brain machine interfaces with multiphoton microscopy

Riichiro Hira

Department of Physiology and Cell Biology, Graduate School of Medical and Dental Sciences, Tokyo Medical and Dental University, Tokyo, Japan

**Correspondence**
Riichiro Hira, M.D., Ph.D.
rhira.phy2@tmd.ac.jp

**Disclosures**
The authors declare no conflicts of interest.

**Acknowledgments**
We thank Drs. S.L. Smith, C.H. Yu, K. Isobe, S. Tsutsumi, and N. Honkura for helpful discussion. This research was supported by JP22wm0525007 from AMED, JP22H02731, JP20K22678, JP21B304, JP21H05134, and JP21H05135 from MEXT/JSPS, Nakatani Foundation, Shimadzu Foundation, Takeda Science Foundation, The Precise Measurement Technology Promotion Foundation, and Research Foundation for Opto-Science and Technology.

**Abstract**

In the field of neuroscience, the importance of constructing closed-loop experimental systems has increased in conjunction with technological advances in measuring and controlling neural activity in live animals. This paper provides an overview of recent technological advances in the field, focusing on closed-loop experimental systems where multiphoton microscopy (the only method capable of recording and controlling targeted population activity of neurons at a single-cell resolution in vivo) works through real-time feedback. Specifically, we present some examples of brain machine interfaces (BMIs) using in vivo two-photon calcium imaging and discuss applications of two-photon optogenetic stimulation and adaptive optics to real-time BMIs. We also consider conditions for realizing future optical BMIs at the synaptic level, and their possible roles in understanding the computational principles of the brain.

*1. Introduction*

In neurophysiological experiments, the experimenter's visual, auditory, and somatosensory perception of the state of the living animals and cells has traditionally been crucial to maintain and observe their physiological functions by adjusting the experimental conditions in real-time. In this sense, closed-loop experimental systems involving the experimenter (i.e., the human operator) appear in various physiological experiments. For example, in patch-clamp recording, a square-wave is given, and its resistance is read by the eye from the oscilloscope or by the ear when recordings are made audible. The resistance readings can then be used to control the manipulator and suction to achieve gigaseal formation. In two-photon imaging, the acquired images are also checked in real-time to guide the visualization of the desired layer and desired cell population. These are processes integrating animals, cells, recording systems, and humans. In a broader sense, obtaining experimental data, creating a new hypothesis through analysis, and conducting a new experiment to test this hypothesis is also a closed experimental system with human intervention. We can ask the question "what is the need for 'human' intervention in such experiments"? It might be because there is no way to automate the pattern recognition performed by human vision and audition, or because it is difficult to control equipment more precisely than through human manipulation. In this respect, current developments in machine learning, including deep learning and control systems, have allowed levels of accuracy outperforming those of humans. Accordingly, various kinds of closed-loop experiments have been realized, such as automated patch-clamp methods assisted by two-photon imaging [1-4] and brain machine interfaces (BMIs) using two-photon imaging [5-10]. In this review, we discuss such closed-loop systems integrating animals, cells, recording and controlling equipment, and machine learning, and operating without human intervention. Such closed-loop systems have the potential to revolutionize experimental speeds and scales by eliminating the need for human involvement in the loop. In particular, we shed light on the use of multiphoton microscopy in closed-loop systems, where it has notably succeeded in achieving both high speed and large-scale operations.

We start by outlining the fundamentals of two-photon calcium imaging and its integration into BMI technology, and also discuss the technical development of two-photon optogenetic stimulation and its real-time applications. The combination of these techniques provides the foundation for all-optical experiments. We address the relationship between intentional control and temporal delay, which must be considered when incorporating an all-optical system into a closed-loop experimental setup. The discussion then moves on to two-photon imaging and control of synaptic strength and activity, which may have potential applications in future BMI research. We also provide an overview of adaptive optics (AO) for performing such synaptic-scale experiments in the deep brain. Finally, we discuss future technologies, including the integration of the brain and artificial intelligence (AI), which can be potentially achieved with these approaches.

## 2. Brain machine interfaces using in vivo multiphoton calcium imaging

### 2.1. Two-photon imaging and BMI

Two-photon calcium imaging is the best existing imaging method for observing in vivo cell morphology and activity in deep brain tissue (< ~1 mm) [11]. In the brain, the number of neurons that can be simultaneously recorded with two-photon calcium imaging exceeds 1 million [12], about two orders of magnitude more neurons than can be recorded with electrophysiological recording using multi-site electrodes such as Neuropixels 2.0 [13]. This very substantial increase in the recording number was made possible by the combination of bright genetically encoded calcium sensors [14-17], large field-of-view (FOV) two-photon microscopes [18-21], and advanced scanning technologies such as reverberation microscopy [12, 22]. BMIs use neural activity to manipulate external devices according to the subject's intentions; however, the neural activity recorded by conventional BMIs is limited to a few hundred neurons at most because of the limits of the electrophysiological recording. Although electrophysiological recording is also currently making progress, with the CMOS process and nanotechnology, the advantages of optical measurement have been clearly established in terms of dense measurement of neurons and the ability to measure without directly penetrating the brain. Furthermore, two-photon optogenetics also uniquely facilitates precise stimulation of targeted cell populations in space and time [23-29], measurement of changes in the membrane potential of individual neurons, even below the threshold of spike generation [30-35], and specific measurement of neurotransmitters such as glutamate, dopamine, and acetylcholine [33, 36-40]. Thus, future BMIs are likely to be optics-based [41]. We will begin by outlining the experiments that have been realized with the current two-photon BMI (2p-BMI).

### 2.2 Examples of 2p-BMI

Hira et al. performed a BMI task in which two-photon imaging with G-CaMP7 was used to convert the activity of a single target neuron into a reward in real time [6]. Mice were trained on a lever-pulling task before this BMI task (Fig. 1a,b). When switched from the lever-pulling task to the BMI task, the same amount of reward was obtained by continuing to pull the lever in the same way if the target neuron was associated with lever-pulling. By contrast, when the target cells were not associated with lever pulling, mice significantly increased reward acquisition by increasing the frequency of calcium transients in the target neurons without lever pulling over a task duration of 15 minutes. Furthermore, analysis of changes in non-target neuron activity recorded by two-photon calcium imaging showed that neurons did not increase their activity in the same way just because they were in the proximity of the target neuron; neurons that were coincidentally synchronously active just before reward acquisition increased their activity over the 15 minutes, whereas neurons that showed activity after the reward was obtained showed a decrease in activity within 15 minutes. This indicates that individual neuron activity was enhanced when the activity was causally related to the reward, and suppressed when the temporal order was reversed. The authors were able to reproduce this bidirectional change in activity by manipulating neural activity with one-photon optogenetics and substituting the timing of the reward (Fig. 1c,d,e). These results suggest that the timing of neural activity relative to the reward plays an important role in the proper control of a BMI.

Clancy et al. used GCaMP6f to perform two-photon calcium imaging to measure the difference in activity between two populations of neurons in layer 2/3 of mouse M1 (primary motor cortex) in real time [5]. They converted the difference into a sound and fed it back to the mice, while at the same time rewarding them when the value exceeded a threshold, thereby encouraging intentional manipulation of neural activity. This training resulted in a change in the activity of the two populations and led to efficient obtainment of the reward. By examining the changes in the activity of the neurons (recorded by two-photon calcium imaging) that were not used for real-time feedback, the authors found that neurons close to those used for control changed their activity in the same way.

Mitani et al. performed two-photon calcium imaging specifically on inhibitory cell types [7]. For example, Cre-dependent expression of calcium sensor enabled parvalbumin-positive neuron-specific recording. They divided the parvalbumin-positive neurons into two populations and gave rewards when the sum of one population activity minus the sum of the other population activity exceeded a threshold value. They repeated this experiment with somatostatin-positive and vasoactive intestinal peptide-positive neurons, and found that while all three neuron types were successfully volitionally controlled, only the parvalbumin-positive neurons were controlled by decreasing the activity, whereas somatostatin and vasoactive intestinal peptide neurons were controlled by increasing the activity. This indicates that in addition to flexible controllability of inhibitory neurons, intentional control differs depending on the subtype. In addition, Mitani et al. published a paper on a module for real-time analysis of two-photon calcium imaging data [42].

Differences in cell-type-dependent intentional regulation are also seen in excitatory neurons. Nuria Vendrell-Llopis and colleagues expressed a calcium sensor in an intratelencephalic neuron-specific or pyramidal tract neuron-specific manner, and used it to control a BMI [10]. The results clearly showed that volitional control was easier for pyramidal tract neurons. Together with the results of Mitani and colleagues [7], these results show that the cortical circuitry involved in volitional control is highly structured by the cell-type.

Prsa et al. performed two-photon calcium imaging using GCaMP6f to measure the activity of a neuron in M1 in real-time for operant conditioning of single neurons [8]. At the same time, other neurons in S1 were stimulated with one-photon optogenetics at an intensity proportional to the activity of the recorded neuron in M1. This feedback was found to significantly increase the efficiency of BMI training, probably because it worked as a conditional reinforcer. This strategy was also successfully used to manipulate a robotic arm.

*2.3. Current status of 2p-BMIs for human applications*
As described above, BMI using two-photon calcium imaging has already contributed to our understanding of the computational principles of the brain [43]. In respect to this work, we can ask the question "what is the situation considering its application in humans for driving a neuroprosthetic device"? Trautmann et al. first demonstrated a 2p-BMI in non-human primates using two-photon calcium imaging in macaques [9]. The macaque cortex is several times thicker than that of mice, and thus imaging the cell body of layer 5 Betz cells is difficult. Therefore, they used adeno-associated virus vectors to express GCaMP6f in dorsal premotor (PMd) or M1 cortices, and performed temporal (13 days, seven sessions) imaging of dendritic activity in layer 1. They demonstrated that dendritic activity can be used for decoding the movement of the arm.

This successful decoding of movement via two-photon imaging in primates is a major step toward its application in humans. Gene transduction in humans for the restoration of vision and hearing in hereditary diseases is already being studied [44]. Gene transfer with virus vectors to the human brain is also being developed because of the need for gene therapy for glioblastoma and other diseases [44-46]. Another important step for human applications of BMI in daily life is miniaturization of the ultrashort pulsed laser, scan optics, and detectors used for two-photon imaging and two-photon stimulation (as discussed below), so that they can be placed on the head. Such miniaturization of the scanning optics is already underway [47-51]; in particular, the latest MINI2P [50] realized three-dimensional scanning using a MEMS (micro-electro-mechanical system) scanner and a MEMS tunable lens. However, no work has yet been done on the miniaturization of the laser and photo-detectors. Major technological innovations will be required in this area in the future.

*2.4. Volitional control and delay time*

Even though the body has many degrees of freedom, the execution of smooth movements requiring the precise control of individual muscles does not require intentional effort. If the mechanism behind this smooth control can be incorporated into BMIs, smooth BMI control with many degrees of freedom and no burden on the subject should be possible. While the brain outputs motor control signals, it also estimates the results of its own movement and predicts the sensory input that will occur after the output. By detecting sensory inputs and correcting for errors between the proprioceptive [52] and/or visual feedback (that occurs after motor output) and the estimated results, the brain can acquire an internal model enabling motor skills [52-55]. Estimation of the consequences of one's own locomotion according to an internal model (forward model) becomes more difficult as the time lag between the generation of the motor output signal produced by the brain and the resulting sensory feedback increases. For example, when balancing upright, the sensory feedback associated with lower limb motor commands is delayed by about 100 ms [56]. Giving sensory feedback with a 200 ms delay makes it difficult to maintain balance in the upright position [57]. Furthermore, tactile sensory feedback for touch according to one's own movements provokes significantly greater ticklishness when delayed by 100 ms compared with actual feedback [58]. The maximum delay for visual feedback to motor output was 230 ms, beyond which subjects perceived misalignment [59]. Thus, although there are some differences between modalities, a rough estimate suggests that sensory feedback within or around 100 ms is important for control via the internal model. This suggests that feedback within 100 ms would also be beneficial for BMI control.

By contrast, in operant conditioning, a delay of up to a few seconds between the behavior to be reinforced and the outcome is usually acceptable. This is probably because the dopamine-dependent plasticity time window of the corticostriatal synapse is about one second [60], and there is a mechanism in the brain for credit assignment to earlier neural activity [61]. Thus, the time window over which reinforcement learning is possible (~1 s) is an order of magnitude larger than the time window over which supervised learning using internal models is possible (~100 ms).

The two-photon calcium imaging described above was conducted at 3–30 frames per second, with a delay of about 30–300 ms. In addition, it takes about 100 ms from neural activity for the calcium concentration to reach a threshold. Considering the delay between real-time motion correction, ROI segmentation, and reward output using a mechanical pump, there should be a delay of 200–500 ms in total. This is shorter than the time available for credit assignment for behavioral reinforcement through operant conditioning, but is longer than the delay available for learning with the internal model. In this case, the 2p-BMI control could be learned to a certain degree by reinforcement learning, but the feedback time should be faster when there are more degrees of freedom to manipulate external devices. To shorten the feedback time, it is necessary to develop sensors with faster kinetics and speedy image processing algorithms for motion correction and segmentation. Recently, fast calcium sensor jGCaMPf was developed, where the delay from the onset of the action potential to half of the peak fluorescence value was about 2 ms [62], which is comparable with that of calcium sensors based on synthetic organic compounds [63]. Real-time image processing is addressed in the next section.

*2.5. Real-time image processing and its speed*
As we noted above, feedback within ~100 ms is needed for control of one's own body movements, with this allowing for a natural application of internal models, rather than feedback within ~1 s, which is the time delay allowing for reinforcement learning. Is it possible to record neuronal activity with two-photon imaging and loop through to compose the appropriate feedback within this time?

There are already several algorithms for registration and ROI segmentation of neural cell bodies in two-photon calcium imaging data [64]. Here, we summarize particularly important papers on real-time use. Mitani and colleagues [42] completed motion correction of 1000 512 × 512-pixel video frames in less than 3 s and showed that such processing can be applied to BMIs [7]. Giovannucci and colleagues [65] developed an algorithm called CaImAn (Calcium Imaging data Analysis) that can be used online with the fast motion correction algorithm NoRMCOrre [66] and ROI segmentation. Bao and colleagues [67] utilized U-Net [68] for this purpose, and successfully extracted active neurons from two-photon calcium imaging data in real-time. CITE-On, a platform for real-time convolutional neural network (CNN) analysis of large FOV two-photon calcium imaging, was also reported [69], and is currently considered to be the best algorithm. CITE-On is much faster than other algorithms because it does not perform real-time correction for motion of less than 4 μm. Even when the number of neurons to be detected is several thousand, the analysis can be completed within 10 ms, making CITE-On suitable for large-scale imaging at 100 fps.

So far, we have assumed imaging of a rectangular region using a resonant scanner and a galvanometer scanner, and have introduced real-time processing of rectangular images with the same FOV. By contrast, if the motion can be handled at the level of the scanning, there should be no need for image processing. The fastest closed-loop scanning system currently available is the RT-3DMC (real-time 3D movement correction) system using two field-programmable gate arrays (FPGAs) [70]. In the study describing this system, instead of scanning rectangular regions, the authors used acousto-optic lens (AOL) 3D scanners to obtain fast random access to the cell bodies of neurons. To compensate for the effect of brain motion, scanner movement was modified by an FPGA according to estimation of the three-dimensional motion by another FPGA, thereby facilitating constant targeting of the cell bodies. The delay time between the brain motion and the control of the AOL was less than 1 ms (395–669 μs). Thus, both the CITE-ON CNN-based fast motion tracking and the RT-3DMC with two FPGAs provide fast and accurate image processing for real-time purposes.

*2.6. Closed-loop multiphoton calcium imaging for purposes other than BMIs*
Studies using BMIs have monitored neural activity and linked it to rewards in order to facilitate learning in the animal brain. In addition to such closed-loop experiments for operant conditioning or BMIs, there are other unique studies using two-photon imaging in real-time, two examples of which we describe below.

One such study used two-photon imaging to automate in vivo patch-clamp recording systems targeting fluorescently labeled neurons. Automation of the patch-clamp method had already been performed for so-called blind patches, but recording against a targeted cell required a human to analyze the image and guide a pipette close to the cell [3]. In two papers published in 2017[1, 4], the images acquired by two-photon imaging were analyzed in real-time to allow the glass pipette to be operated with a manipulator to increase or decrease the pipette's internal pressure at the right time, thereby achieving full automation. The success rate reached by this system was equivalent to that of a skilled human. Such automation can substantially reduce the stress on the experimenter when applied to multi-patches [2]. It may also be applicable to experimental systems that repeatedly measure the RNA levels of the same cell, systems that have been developed in recent years with the increasing use of single-cell RNA-seq technology [71]. Indeed, it is feasible to electroporate a plasmid in the brain of a live animal and then perform patch-clamp recording on the very same cell [72]. Without automation, it would be impossible to perform multiple cellular operations that repeatedly access the internal chemical environment of a living cell.

The second application utilized a closed loop to examine the receptive fields of cellular activity in the visual cortex [73, 74]. Two-photon calcium imaging was used to record the firing activity of neurons in the visual cortex while the animal was exposed to various stimuli. The deep neural network (DNN), trained to be a 'digital twin' mimicking the visual cortex, searched for the complex receptive fields of recorded neurons from pairings of visual stimuli and corresponding neuron activities. The images that were predicted to enhance the activity of neurons were then given back to the animal to see if the activity was actually enhanced. By repeating this loop, the authors were able to determine the most exciting inputs (MEIs) [73]. Moreover, the closed-loop also determined the diverse exciting inputs (DEIs), which are sets of dissimilar visual stimuli that together enhance the activity of neurons in the visual cortex [74]. It is particularly noteworthy that the DEIs revealed, for the first time, a configuration of receptive fields that could be useful for object segmentation according to texture boundaries [74]. As the authors pointed out, such a finding would be difficult to make without a closed-loop experiment. In the future, this method should contribute to our understanding of the complex response properties of neurons in the association cortices, integrating not only visual, but also auditory and somatosensory information.

*2.7. Sensory feedback for comfortable BMI control*
Even after subjects learn how to control external devices through a BMI, they may need to tune it repeatedly because of deterioration in the decoder performance due to neuronal representational drift [75]. Alternatively, the decoder itself, rather than the subject, may need to change from moment to moment, even when the recording is stable [76, 77]. In both cases, even if the volitional control appears to be effortless, it may require the subject to pay a tuning effort. On the other hand, stable low-dimensional representations can be obtained over several years when a sufficiently large number of neurons are recorded [78]. BMI control using such stable low-dimensional dynamics would be very tractable for subjects [79]. It is important to develop algorithms to efficiently extract latent dynamics from large-scale calcium imaging data [80]. Alternatively, if the prediction error signal in the internal model could be extracted directly from the recorded neural activity, it might be possible to use it to directly modify the decoder.

If the BMI is also to be used to complement and extend sensory functions in addition to manipulation of external devices, the photostimulation parameters of the sensory cortex and other neuronal populations must be determined in real-time. In this context, pioneering work by Prsa and colleagues [8] varied the intensity of one-photon stimulation on the sensory cortex in real-time. In another study, an electrophysiological study using a BMI with sensory feedback with a delay of less than 100 ms allowed recognition of object texture mimicking cutaneous sensations associated with motion [81].

If such methods could be extended to provide virtual sensory feedback to targeted neural populations in the cerebral cortex (for example, primary somatosensory cortex) at the right time, further improvement in BMI control should be possible. For this purpose, it is necessary to select the next population to be stimulated in real time according to the recorded neuronal activity. Furthermore, a delay of about 100 ms or less for the entire process needs to be achieved for natural feedback control with an internal model. We will see in the next section how we can quickly target and photostimulate a population of neurons.

## 3. Two-photon photostimulation of neurons
### 3.1. Methods for two-photon photostimulation

An important step in the further development of BMIs is the complementation and extension of artificial sensory feedback [81]. Since sensory feedback is crucial for precise movements [82], the precise manipulation of devices with a high degree of freedom by a BMI requires sophisticated ways to provide appropriate feedback to the brain. Visual recognition of a cursor on a computer or a robotic arm can provide feedback on the results of manipulation, but this is less precise and more delayed than somatosensory feedback.

Restoring vision through the introduction of opsin into retinal neurons is one example of mimicking sensory input by optogenetics. The retina is composed of thin layers of cells that can be stimulated with high resolution, even by one-photon stimulation; the way the eye is structurally evolved facilitates this. However, when such a method is applied directly to the brain, it is necessary to properly illuminate the population of neurons arranged three dimensionally in the deep brain. One-photon stimulation is useless in the case of L2/3 (~ 200 μm) or deeper because of strong scattering, and two-photon optogenetics is necessary to stimulate a large number of targeted neurons.

Two-photon photostimulation has advanced remarkably over the past decade, with the development of both opsin and optics, and further progress is expected in the future. Two-photon excitation of opsins such as channel rhodopsin 2 (ChR2), like that of ordinary fluorescent molecules, is evoked by an ultrashort pulsed laser with a wavelength about twice the one-photon absorption wavelength. Whereas the time required for fluorescent molecules to return to the ground state after excitation is on the order of ns, that of the state changes of ChR2 is on the order of μm to ms. In addition, the peak power required for two-photon excitation seems to be higher for opsin than for fluorescent molecules. Therefore, unlike fluorescent molecules, two-photon excitation of opsin uses lasers with higher peak power and lower repetition rate [27, 28].

Opsins are expressed in two dimensions within the cell membrane, making excitation inefficient compared with fluorescent molecules distributed in three dimensions within the cytoplasm. For this reason, the early studies used a galvanometer scanner to scan spirally around the cell body to efficiently stimulate membrane opsins. Later, a holographic method using a spatial light modulator (SLM) was developed to simultaneously stimulate multiple neurons, and temporal focusing [83] was added to increase the resolution in the depth direction of holographic focus [84, 85]. Furthermore, the 3D-SHOT technique was developed, which involves putting diffraction gratings before SLMs to allow efficient three-dimensional photostimulation [23, 25]. Currently, the most efficient opsin for this purpose would be either modified versions of ChRmine [86, 87] or modified versions of ChroME [88]. A combination of 3D-SHOT and ChroME2s realized targeted photostimulation of more than 600 cells per second [89]. 3D-SHOT can be used with large FOV two-photon microscopy [90], which should enable the control of more neural activities.

*3.2. Speed of two-photon photostimulation*
How many neurons can be stimulated per second with the current level of technology? Let us assume that each cell needs to be stimulated with a laser of about 10 mW for about 10 ms. The upper limit of total laser power would be about 200 mW due to the problem of overheating caused by water absorption [91]. Thus, the limit would be 100 patterns of 20 populations, a total of 2000 neurons/s. The upper limit of laser power at which a large FOV microscope can be used simultaneously may be around 400 mW [12], and the laser power for single neuron stimulation for 10 ms could possibly be shortened several times, but at current levels, simultaneous stimulation of 10,000 neurons/s would be difficult. This is two orders of magnitude smaller than the number that can be simultaneously imaged (1 million neurons [12]).

To stimulate a targeted population as simultaneously as possible, it is necessary to create a computer-generated holography (CGH) pattern for the SLM on the fly. The GS algorithm [92], which is still the one most commonly used for holographic stimulation, requires iterative calculations and takes 200–300 ms to construct a single pattern. DeepCGH [93] succeeded in creating an arbitrary CGH pattern with high accuracy through unsupervised deep learning of images that are close to the target image among those that can be constructed using an SLM. The trained CNN was able to create SLM patterns an order of magnitude faster than the GS algorithm, at 8.7 ms per image. The computational speed of CNNs is expected to increase further with the development of GPUs.

Various SLMs have been developed in recent years. The fastest liquid crystal on silicon (LCoS)-SLM currently available is from Meadowlark Optics, with a claimed refresh rate of 2 kHz for a 1024 × 1024 resolution. In reality, it takes 1.7 ms to stabilize the liquid crystal at the 1064 nm used for two-photon stimulation, so the stimulation must be turned off during this time. Nevertheless, this latency is shorter than that of SLM pattern formation with DeepCGH. Currently, SLMs have achieved 10 kHz for 100 × 100 pixels, and faster SLMs with higher resolution will be developed. Since spatiotemporal focusing does not use the entire two-dimensional surface of the SLM, multiple patterns can be placed at multiple locations on the SLM, and the scanner can select one of the patterns in order to change the holographic pattern at high speed [94]. Thus, improvements in the spatial resolution of the SLM would also improve the speed of photosimulation.

Currently available SLM holography generally produces a 20 μm-diameter spherical excitation spot for a neuron when 10 or more neurons needed to be photoactivated simultaneity. In reality, this is sufficient to excite molecules on the membrane of a neuron with a diameter of about 10 μm, but it would be more efficient if the SLM could produce a spherical pattern that precisely matches the shape of the cell membrane as succeeded in the case of a single neuron [95]. This issue may be resolved as the resolution of SLM increases and/or methods for the spatiotemporal focusing are improved. In addition, localized expression of opsin near the cell body will become more important as the number of stimulated cells increases [96-98].

A combination of imaging and stimulation with two-photon microscopy, i.e., an all-optical experiment in which activity is monitored by two-photon imaging and simultaneously controlled by two-photon stimulation of opsin, has been realized. What is possible when we can record and stimulate at the same time? In the case of whole-cell patch-clamp recording for monitoring and stimulating the cell, one can hold the membrane potential according to the negative feedback of the operational amplifier. A similar experiment can now be performed noninvasively with a two-photon microscope [27]. More experiments would be realized with an all-optical two-photon microscope contributing to a BMI, a point we discuss in the next section.

*4. Interaction of brain and machine*

Techniques that read and process neuronal population activity and then utilize it to stimulate the activity of another neuronal population can advance the integration of the computer and the brain [99, 100]. For example, in a pioneering study, single neuron activity was recorded in the motor cortex of a monkey, and immediately after detection of a spike, a neural population in another region was electrically stimulated [101]. This resulted in a strong functional coupling between the neuron used for triggering and the stimulated population. This artificial functional connection was created only when the time between recording and stimulation was less than 50 ms. The authors argued that a multi-synaptic spike-time dependent plasticity (STDP) [102] was involved in this functional connection. A similar experiment was performed with two-photon microscopy [27]. While performing two-photon recording of the activity of a group of neurons that simultaneously expressed a calcium indicator (GCaMP6f) and an opsin (C1V1), the researchers photostimulated tens of pre-determined neurons by two-photon laser immediately after the calcium transient of a trigger neuron occurred. They showed that the increase in the activity of the stimulated population was significantly greater when it was triggered by the trigger neuron than when it was stimulated randomly. However, in this system, the time between the firing of the trigger neuron and the onset of stimulation may not be less than 50 ms (for details, please see Supplementary Figure 6 of Zhand et al 2018).

When one can pre-determine the populations of neurons for photostimulation, the SLM pattern can be obtained before the experiment. In this case, one does not need to construct a new SLM pattern on the fly. However, if one were to select the neurons for photostimulation based on the recording and then photostimulate them by determining the SLM pattern in real-time, application of this close-loop experiment could be further expanded. For example, we would be able to establish communication between the brain and AI. Although currently not yet successfully performed, it is in principle possible to utilize ongoing neural population activity to drive an artificial spiking neural network (SNN) and feedback the weighted sum of the activity of the SNN units through two-photon stimulation in real time. If such a system could learn to provide appropriate feedback reflecting the brain state in real-time, it could suppress epileptic seizures and replace deep brain stimulation for treating Parkinson's disease [41]. Moreover, it could also extend the function of the brain, as foretold in science fiction novels.

## 5. Closed-loop experiments at the synaptic level
### 5.1. Synapse and learning

So far, our discussion has focused on how to record and control the neuronal cell body, but not the synapse. Since the brain is a system that learns by plastic changes in the strength of synaptic connections, direct measurement of synaptic strengths and the ability to modify them will contribute to recovery or augmentation of brain functions in the future. For example, as in the error backpropagation of deep learning [103], it would be possible to optimize an arbitrary objective function in the local circuit of the brain by estimating the contribution of individual synapses to this objective function and manipulating them according to the contribution. Even if the brain is not a circuit where exact error backpropagation is computed, a credit assignment for each synapse would still be possible with locally available information that each synapse can access [104-106]. In this way, by hijacking biological learning rules for optimization of the objective function by synaptic plasticity, the brain may be able to "directly learn" desired functions. Thus, technologies to measure and manipulate synapses in local circuits will take brain-AI integrated systems to the next level.

### 5.2 Optical measurement and control of synaptic strength and synaptic activity

In the cerebral cortex, the weight of excitatory synaptic connections is proportional to the volume of the dendritic spine, a post-synaptic structure that receives synaptic inputs from axonal boutons [107]. Therefore, if fluorescent proteins are expressed in the spine and the volume of the spine can be measured by two-photon imaging or a super-resolution stimulated emission depletion (STED) microscope based on 2P excitation [108], the strength of individual synapses can be estimated. In addition, by using fluorescent proteins such as iGluSnFR3 [37], which can directly visualize glutamate (the neurotransmitter of excitatory synapses), the amount of transmitter released between synaptic clefts can be estimated. Calcium imaging of the spines can also be used to estimate the influx of calcium ions due to synaptic inputs [109]. Recently, direct observation of spines by two-photon voltage imaging was realized [35]. Many technologies, such as multiple beams strategy [110], tomography [33], laser multiplexing with a FACED module [32], and the 16 PMTs strategy [30], could contribute to high-speed scanning for tracking of changes in membrane potential over a larger scale. Thus, the volume, glutamate, calcium, and membrane voltage of single spines are now all observable in vivo. In addition to recording the features of post-synaptic spines, direct axon imaging and photostimulation of pre-synaptic neurons would be useful for identifying pre-synaptic partners for each synapse.

Is it possible to use photostimulation to artificially impose inputs on targeted single synapses? It is already possible to target two-photon photostimulation to spines using C1V1 opsin [24]. However, it should be noted that although the input depends on the amount of C1V1 expression on the spines, it is independent of the synaptic strength. Optical stimulation of axons would be useful to induce natural synaptic transmission. It was shown that while one-photon stimulation of an axon readily induces action potentials, two-photon stimulation is less likely to do so [26]. Moreover, once an action potential is triggered, synaptic inputs can occur anywhere on the same axon, making it difficult to localize them to a targeted synapse. To stimulate only the targeted pre-synapses, stimulation must be applied without triggering action potentials. The sCRACM (subcellular ChR2-assisted circuit mapping) method involves photostimulating exocytosis only at the photostimulated pre-synaptic area in the presence of tetrodotoxin (TTX) [111]. However, this method cannot be used in vivo because of the TTX. Stimulation of pre-synaptic terminals without triggering action potentials requires direct access to the molecular mechanisms that would occur after pre-synaptic depolarization. This may be possible by directly controlling the exocytosis of pre-synaptic terminals. Light-sensitive Gi proteins such as eOPN3 [112] and PPO [113] can suppress synaptic transmission when expressed pre-synaptically, and are potentially usable with two-photon excitation, although they may cause effects via activation of Kir channels at pre-synapses. In this respect, opto-SynC [114], which uses light-evoked homo-oligomerization of cryptochrome CRY2 to cluster synaptic vesicles and inhibit exocytosis, would be completely independent of membrane potential. If the opposite is also possible (i.e., photoactivation of Gs and Gq proteins (opto-XR [115]), photoactivation of cAMP (bPAC) [116], or light-evoked clustering of synaptic proteins) and exocytosis can be facilitated with two-photon targeted photostimulation, it would be possible to stimulate targeted synapses.

In this respect, can we artificially alter the strength of synaptic connections? The molecular biology of synaptic plasticity is one of the best understood areas of neurophysiology, and we know that a variety of molecules are involved in the synaptic strength and its plasticity. Optogenetic regulation of Ras [117] and CaMK2 [118] confined to the spine, regulation of the endocytosis of AMPA receptors in the post-synapse [119], and control of pre-synaptic cAMP [120] have already been successfully used to manipulate plasticity. Thus, it is possible to regulate plasticity up or down by two-photon stimulation of a targeted spine, as in the various intracellular compartments [121]. Such technologies could contribute to clinical medicine in the future by restoring or enhancing brain function or erasing unwanted memories [122].

*5.3. Adaptive optics for synapse-level BMIs*
Spines are an order of magnitude smaller than neurons. Accordingly, two-photon microscopy only allows clear in vivo imaging of spines in layer 1. To visualize synaptic structures with diffraction-limited resolution in the deeper layer of the brain, AO should be incorporated into two-photon microscopy [123-125]. AO can optimize the effectiveness of photons in terms of the excitation of molecules and enhance the resolution of imaging by modifying the wavefront of the laser. The challenge for AO is to obtain information on how the wavefront is distorted inside the brain tissue and compute it backward. Closed-loop control is also required for this computing process.

The simplest case is to perform two-photon imaging and evaluate the excitation efficiency by averaging the intensity of the fluorescence image while changing the parameters of the deformable mirror (DM) [126], which is also used in large FOV two-photon microscopy [18]. Zernicke modes up to the 15th order have been optimized with this strategy, and the method has also been applied to three-photon imaging [127]. The advantage of this simple but powerful method is that it does not require any equipment or devices other than a DM.

The most straightforward way to efficiently perform AO is to directly observe the wavefront with "a guide star". Wang et al. injected indocyanine green directly into the brain and used it as a guide star [128]. Liu et al. introduced Cy5.5-dextran into blood vessels by retro-orbital injection and used its fluorescence as a guide star when it traveled through the blood vessels and filled the brain [129]. The fluorescence from the guide star entered the optical pathway from the brain in the opposite direction to the excitation laser and reached the Shack-Hartmann wavefront sensor (SHWS) at the pupil plane, with this sensor being able to directly measure the wavefront. By optimizing the parameters of the DM according to the wavefront information, they were able to modify the wavefront of the excitation laser and dramatically increase the excitation efficiency of fluorescent proteins. As a result, Wang et al. succeeded in clear calcium imaging at a depth of 700 μm from the brain surface, and Liu et al. succeeded in observing dendritic activity and glutamate release from axons in layer 5b of the mouse cortex.

Na Ji and colleagues successfully performed high-resolution wavefront modification using SLM without a guide star [130]. They first divided the pupil plane into segments and performed laser excitation through each compartment. They modified the wavefront of the laser with SLM conjugated to the pupil plane in each segment so that the excitation by the laser through each segment occurred at the same location. This cycle was repeated several times, and once the optimization of the wavefront of the laser was completed in all segments, a high-resolution image was successfully obtained using all segments. A parallelized version of this method was also developed to speed up the process [131]. This enabled observation of L4 axons and synaptic structures and activities [132]. This pupil segmentation strategy was also used for three-photon excitation [133].

The three methods described above can correct for aberration but not for finer μm-scale scattering. To deal with scattering it is necessary to compensate for large degrees of freedom, including high Zernicke mode coefficients. A method making this possible is conjugate AO [134, 135], which works by placing the SLM or DM on a plane that is conjugated to the layer where scattering is most likely to occur, instead of the pupil plane. To perform AO with two or three photons, the electromagnetic point-spread function (PSF) can be measured using interferometric focus sensing methods [136-138]. These methods have successfully optimized more than 1000 independent Zernicke modes [139], allowing two-photon imaging of small structures such as dendrites and spines in deep brain tissue through thinned skulls.

Although conjugate AO is an order of magnitude better than other methods as far as the accuracy of correction and FOV area, it requires re-measurement when the FOV is shifted more than 100 μm. Since the optimization process takes at least a few seconds, the parameters for each FOV should be pre-determined for large FOV imaging. Supposing that the parameters need to be changed every 100 μm, if a 1 mm FOV is scanned with a resonant scanner at 8 kHz, the parameters must switch at more than 160 kHz. This is well beyond the current technological limits of SLMs or DMs. If there is a large aberration in a single layer, as in a thinned-skull condition, it is sufficient to place an SLM or DM on the conjugate plane of that layer. However, if the aberration that we want to correct is distributed three dimensionally, as in the case of an open skull condition, conjugate AO on a single layer will not be sufficient. Therefore, correction on at least multiple planes (multiconjugate AO, [124, 140]) or SLM with a 3D structure will be necessary in the future [123]. Thus, development of SLMs or DMs is essential for the application of conjugate AO for high-resolution two-photon imaging in a large 3D volume.

We have seen four ways to perform AO. Since the aberrations of brain tissue are considered to be stable across the imaging time (~1 hour), it should be possible to acquire wavefront correction information for each small FOV prior to performing calcium imaging. The fact that real-time acquisition of correction parameters is not necessary is good news for imaging of large volumes in a closed-loop system. However, we also pointed out that conjugate AO, which is necessary for deep brain imaging at the synapse level, requires faster switching of correction parameters when the FOV is large, which limits the feedback time. These issues may be addressed by improving scanning methods and developing faster SLMs and more flexible DMs.

After outlining the recording and control of synaptic strength and synaptic activity using two-photon microscopy, we discussed AO techniques that improve the resolution of two-photon excitation to the synaptic level in deep brain tissue. Integrating these techniques will enable experiments on large numbers of synapses in vivo. We have pointed out that synaptic control based on the identification of synaptic circuits, if possible, has the potential to optimize local circuits for a given objective function like artificial neural networks. Needless to say, we need to advance our understanding of how synaptic plasticity realizes sophisticated brain functions for this purpose. Research on the architecture and learning rules of artificial neural networks and their implementation in the biological brain circuit (so-called "biological plausibility") will become increasingly important [105, 141].

*6. Multiple learning rules and distinct feedback control speeds*
According to Doya [142], different circuits of the brain are responsible for unsupervised learning, supervised learning, and reinforcement learning. As mentioned in the first half of this article, supervised learning and reinforcement learning can be learned with feedback times of ~100 ms and ~1 s, respectively. STDP (spike-timing dependent plasticity) leads to self-organization and can be considered an implementation of unsupervised learning. In STDP, pre-synaptic and post-synaptic action potentials must be controlled to within the order of ~10 ms or less, which suggests that the timescale for unsupervised learning is 10 ms. Therefore, closed-loop experimental systems with time delays of 10 ms, 100 ms, and 1 s are considered essential conditions for the external control of unsupervised learning, supervised learning, and reinforcement learning, respectively (Fig. 2a). Note that although each of these molecular-level mechanisms occurs on a fast timescale, the time delay required for external control differs among them.

The ability to provide feedback control faster than any of the 10 ms, 100 ms, or 1 s thresholds determines what kind of learning rules the closed-loop experiment can implement. This indicates that the ability of a two-photon microscope to record and control the activity of neuronal populations depends critically on the feedback speed that these thresholds can exceed (Fig. 2b). The elements of this closed-loop experimental system include fluorescent molecules, scanning schemes, real-time image processing, AO, the selection of stimulated neurons, CGH, and SLM control, and each must work together in an integrated manner.

For closed-loop experimental systems to reproduce the computational principles of neural circuits, and to create more effective BMIs, it is essential to further increase the number of neurons that can be recorded and controlled in real-time, in addition to the speed. Figure 2c briefly outlines prospects, including technological developments. Considering the development of this field over the last 20 years, it is likely that there will be further unforeseen technological developments in the near future. The recent spectacular developments in multiphoton microscopy and optogenetics provide us with an optimism that the technological problems explicitly illustrated will be solved.

## 7. Conclusion

In this paper, we introduced multiphoton imaging via closed-loop systems and two-photon optogenetics and discussed its potential to significantly advance BMI technologies. Compared with electrophysiological techniques, the advantages of multiphoton microscopy are now clear at the experimental level. However, application of this technology in humans will require major technological innovation, as well as careful discussion and consideration of the ethical issues [143, 144]. Nevertheless, experimental systems that record and control brain dynamics in real time will undoubtedly be key to solving the mysteries of the brain. Such experimental research contributes to a mechanistic understanding of the brain, and we expect to see an acceleration in the applications of it to human diseases in the future.

*8. Figures*

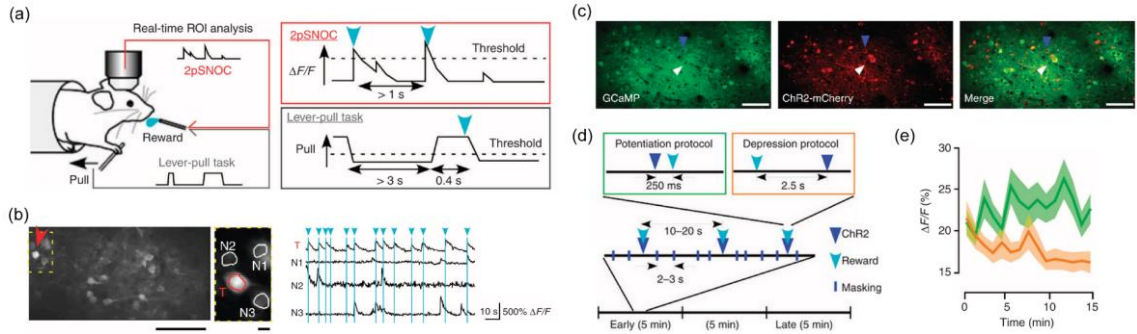

**Figure 1. 2p-BMI**

(a) Head-fixed mice were trained to perform a lever pull task and a single-cell operant conditioning task. (b) ROI analysis was performed in real time by two-photon calcium imaging. The mouse was rewarded by the activity of a target neuron, which leads to elevation of activity of the target neuron. (c) G-CaMP7 Densely expressed G-CaMP7 and sparsely expressed ChR2. (d) Photostimulation was applied 250 ms before or 2.5 s after the reward. (e) In (d), the activity of surrounding neurons was elevated when the stimulus came before the reward, while was decreased when the stimulus came after the reward, which replicated the changes in neuronal activity during single-cell operant conditioning (adapted from Ref. [6] ).

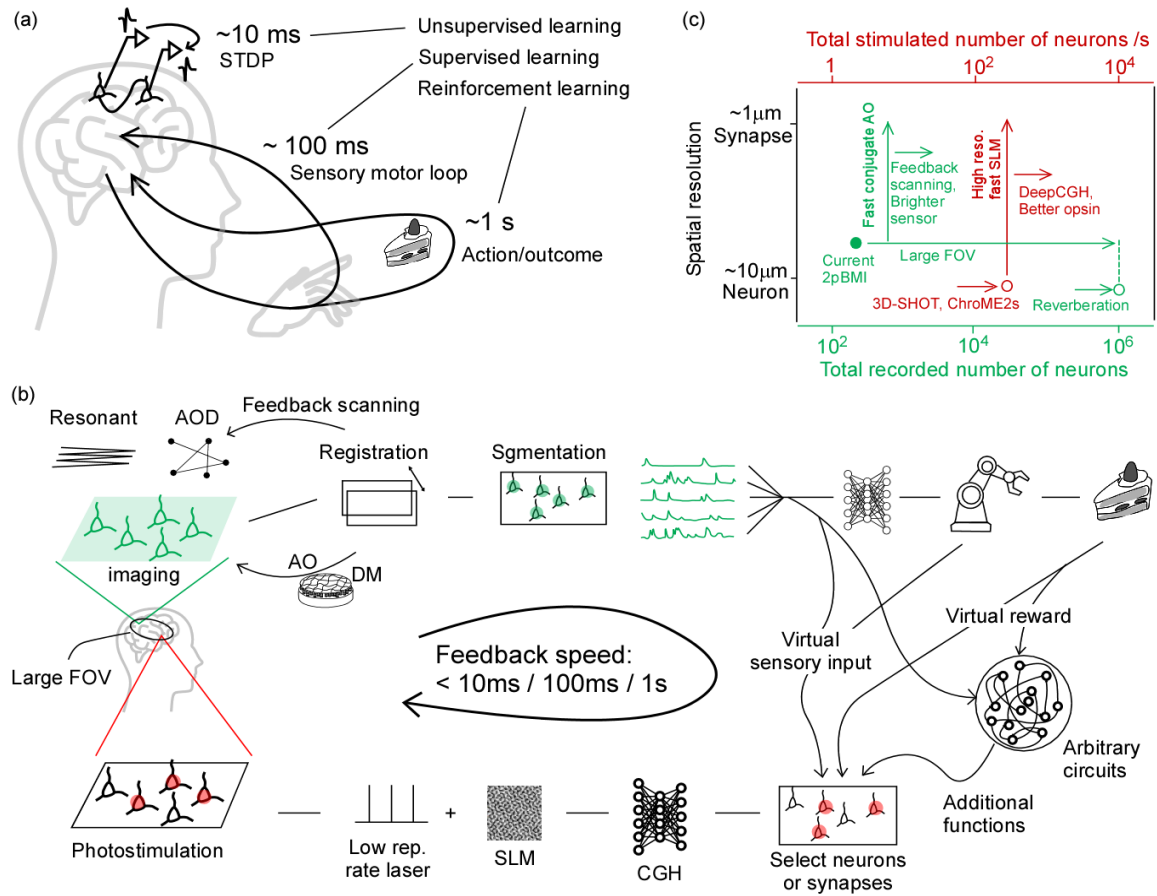

**Figure 2. Natural and artificial closed-loops.**

(a) Three learning rules [142] and their time scales. Unsupervised learning via STDP (spike-timing dependent plasticity), supervised learning via sensory-motor feedback, and reinforcement learning via dopamine-dependent plasticity have distinct feedback times with about an order of magnitude difference. (b) BMI using a combination of two-photon imaging (upper panel) and two-photon photostimulation (lower panel). It is thought that real-time processing within 10 ms, 100 ms, or 1 s will allow us to see the recovery and enhancement of different levels of brain functions. (c) Development of 2pBMI through a combination of two-photon imaging (green) and two-photon photostimulation (red). The horizontal axis (green) is the number of total recorded neurons during the 2pBMI experiment. The filled green is based on Hira et al., 2015, but other 2pBMI studies are not much different. The horizontal axis (red) is the number of neurons per second of the two-photon photostimulation. The vertical axis shows the spatial resolution. The open circle is not the one used for 2pBMI, but is that of the experiment that can be considered the current state-of-the-art. Each arrow indicates the possible path of development that has been or will be considered.